\documentclass[final,5p,times,twocolumn,sort&compress]{elsarticle}

\usepackage{amssymb}
\usepackage{graphicx}
\usepackage{subfigure}
\usepackage{epstopdf}
\usepackage{multirow}
\usepackage{tabularx}
\usepackage{color}
\usepackage{amssymb}
\usepackage{bm}

\usepackage[pdfstartview=FitH, colorlinks,
linkcolor={red!60!black!},
anchorcolor={green!80!black!},
urlcolor={blue!80!black!},
citecolor={blue!50!black!}]{hyperref}

\journal{XXXX}

\begin{document}

\begin{frontmatter}

\title{Two-dimensional quadratic double Weyl semimetal}

\author{Xinlei Zhao$^{1}$}
\author{Fengjie Ma$^{1}$\corref{cor1}}
\author{Peng-Jie Guo$^{2}$\corref{cor2}}
\author{Zhong-Yi Lu$^{3}$}

\address{$^{1}$The Center for Advanced Quantum Studies and Department of Physics, Beijing Normal University, Beijing 100875, China}
\address{$^{2}$Songshan Lake Materials Laboratory, Dongguan, Guangdong 523808, China}
\address{$^{3}$Department of Physics, Renmin University of China, Beijing 100872, China}

\cortext[cor1]{fengjie.ma@bnu.edu.cn}
\cortext[cor2]{pjguo@iphy.ac.cn}

\begin{abstract}
Unconventional Weyl semimetals have attracted intensive research interest in condensed matter physics and materials science, but they are very rare in two dimensions. In this work, based on symmetry analysis and the first-principles electronic structure calculations, we predict that the Si/Bi van der Waals heterostructure is a two-dimensional unconventional quadratic double Weyl semimetal with strong spin-orbit coupling (SOC). Although unprotected by the $C_{3v}$ double group symmetry of the heterostructure, the two-dimensional quadratic double Weyl semimetal is stable for compressive strains up to 6.64\%. The system transforms into a trivial semimetal with further increasing strain, where the phase boundary is a two-dimensional triple degenerate semimetal state. Furthermore, the Kane-Mele tight-binding model calculations show that the quadratic double Weyl phase is derived from the competition between the Rashba SOC and the proximity-effect-enhanced intrinsic SOC. On the other hand, by breaking mirror symmetry, the quadratic double Weyl semimetal transforms into a quantum spin Hall insulator as well as a quantum valley Hall insulator phase. Thus, the Si/Bi heterostructure is an excellent platform for studying the exotic physics of two-dimensional double Weyl semimetal and other novel topological phases.
\end{abstract}


\end{frontmatter}

\section{Introduction}

Weyl semimetals have been intensively studied owing to their many exotic properties, such as chiral anomaly, chiral zero sound, and topologically protected Fermi arc surface states \cite{PhysRevB.83.205101, PhysRevLett.107.186806, PhysRevX.5.011029, Weng_2016,RN556, PhysRevX.5.031023, RN1361, RN1290, PhysRevX.9.021053, PhysRevX.9.031036}. Conventional Weyl nodes have linear dispersion and carry a chiral topological charge $C$=$\pm$1, which do not need any symmetry protection except for discrete translation symmetry. Unconventional Weyl nodes, on the other hand, contain double, triple, and quadruple Weyl nodes and they are protected by certain crystal symmetry, for example, $C_{4}$ or $C_{6}$. Moreover, the double and triple Weyl nodes have quadratic and cubic dispersions and carry chiral topological charges $C$=$\pm$2 and $C$=$\pm$3, respectively \cite{PhysRevLett.107.186806, RN1031, PhysRevB.101.205134}, while the quadruple Weyl nodes carry a chiral topological charge $C$=$\pm$4 with cubic and quadratic dispersions along the [111] direction and other directions, respectively \cite{PhysRevB.102.125148}. Unconventional Weyl semimetals host more topologically protected Fermi arc surface states than conventional ones \cite{PhysRevLett.107.186806, RN1031, PhysRevB.101.205134, PhysRevB.102.125148}.

Currently, the unconventional Weyl semimetals studied are basically three-dimensional. A natural question that arises is whether unconventional Weyl nodes exist in two dimensions. In three dimensions, a Weyl node is surrounded by a two-dimensional closed Fermi surface, so that a chiral topological charge can be defined and calculated. In comparison, in two dimensions, a Weyl node is surrounded by a one-dimensional closed Fermi ring, leading to an ill-defined chiral topological charge. Nevertheless, it is interesting that two-dimensional Weyl nodes can carry $\pm\pi$ or $\pm$2$\pi$ Berry phases and correspondingly have linear or quadratic dispersions \cite{PhysRevLett.102.096801, PhysRevB.92.041404, PhysRevB.78.245122, PhysRevLett.103.046811}. They are similar to the three-dimensional conventional Weyl and unconventional double Weyl nodes, and hence, we here term them as two-dimensional conventional Weyl node and unconventional double Weyl node, respectively. The two-dimensional conventional Weyl semimetals have been well studied \cite{PhysRevLett.116.116803, PhysRevB.94.134513, PhysRevB.100.064408, RN1365, MENG2021148318,PhysRevB.102.075133}, but the two-dimensional unconventional double Weyl semimetals are very rare. Very recently, the first spinful two-dimensional double Weyl semimetal FeB$_2$ is predicted in theory and the Chern number $C$=2 quantum anomaly Hall effect may be fulfilled in FeB$_2$ \cite{wu2021unprotected}. However, due to a weak spin-orbit coupling (SOC), FeB$_2$ is disadvantageous for studying the exotic physical properties of two-dimensional double Weyl semimetals. Novel two-dimensional double Weyl semimetals with strong SOC are therefore highly desired.

In this work, we predict that the Si/Bi van der Waals heterostructure is not only a two-dimensional unconventional quadratic double Weyl semimetal with strong SOC, but also an excellent platform to realize a variety of novel topological phases. A two-dimensional triple degenerate semimetal can be obtained by applying compressive strain, as the phase boundary between the quadratic double Weyl and trivial semimetals. Furthermore, a quantum spin Hall insulator as well as a quantum valley Hall insulator phase can be achieved by breaking the mirror symmetry. As described by the Kane-Mele tight-binding model calculations, both the quadratic double Weyl node and the topological phase transition originate from the competition between the strong Rashba SOC and the proximity-effect-enhanced intrinsic SOC. The Si/Bi van der Waals heterostructure is an ideal arena for exploring various topological phenomena and associated physics.

\begin{figure}
\centering
\includegraphics[width=8.5cm]{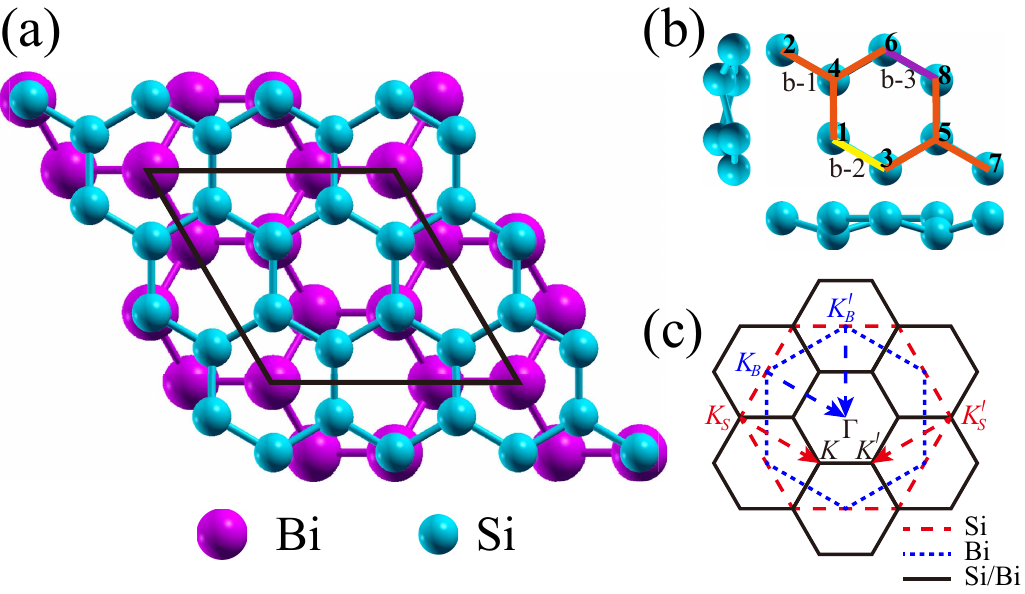}
\caption{\label{crystal}(a) Schematic view of the Si/Bi heterostructure. The black rhombus represents the unit cell used in the calculations. (b) Top and side views of Si atoms in the unit cell. A ship-like warp structure is formed, in which the two sunken Si atoms are labeled by 4 and 5. (c) Schematic of the Brillouin zones of the Si (red), Bi (blue), and Si/Bi heterostructure (black).}
\end{figure}

\section{Computational Details}

In our calculations, the plane-wave basis based method and Quantum-ESPRESSO software package were used \cite{QE2009, QE2017}. We adopted the generalized gradient approximation of Perdew-Burke-Ernzerhof formula for the exchange-correlation potentials \cite{perdew1996generalized}. The ultrasoft pseudopotentials were employed to model the electron-ion interactions \cite{vanderbilt1990soft}. In order to obtain accurate Si-Bi distances, van der Waals interaction was included by using the non-local functionals method \cite{VDWdf}. A 20 $\AA$ vacuum layer was used to avoid the residual interaction between adjacent layers. After full convergence tests, the kinetic energy cutoff for wavefunctions and charge densities were chosen to be 50 and 400 Ry, respectively. In the calculations, all structural geometries were fully optimized to achieve the minimum energy. The tight-binding methods by the combination of Wannier90 \cite{mostofi2008wannier90} and WannierTools \cite{WU2017} software packages based on the Green$'$s function method were adopted to study the edge states.

\section{Si-Bi heterostructure}

The Si/Bi heterostructure was constructed by combining a $\sqrt{3}\times\sqrt{3}$-R30$^{\circ}$ supercell of a single bilayer bismuthene and a 2$\times$2 supercell of planar silicene, both of which have been synthesized experimentally \cite{stpniak2019planar, BLBiPRL,BLBiMiao2758,BLBiWang2013}, as shown in Fig. \ref{crystal}(a). Based on the experimental structures, supercells of both layers were created in commensurate with a common lattice, and the optimized supercell lattice constant was $\sim$7.78 $\AA$. The separation between Si and Bi layers is about 3.06 $\AA$, indicating a van der Waals interaction regime. In the heterostructure, a ship-like warp structure of Si atoms is formed, in which the two Si atoms labeled respectively as 4 and 5 in Fig. \ref{crystal}(b) are relaxed more along the perpendicular $z$ direction than the other six Si atoms. There are three different lengths of Si-Si chemical bonds formed, as labeled respectively by $b$-$1$, $b$-$2$, and $b$-$3$ in Fig. \ref{crystal}(b). The difference between the bond-types $b$-$2$ and $b$-$3$ is very small, both of whom are a little longer than the bond-type $b$-$1$. The overall relaxed structure is mirror-symmetrical along the cell-axis and the short diagonal direction of the calculation cell. Thus the Si/Bi heterostructure has $C_{3v}$ point group symmetry. The corresponding Brillouin zone of the Si/Bi heterostructure along with its high-symmetry k-point folding relationship with Brillouin zones of silicene and bilayer bismuthene are shown in Fig. \ref{crystal}(c).

\begin{figure}
\centering
\includegraphics[width=8.5cm]{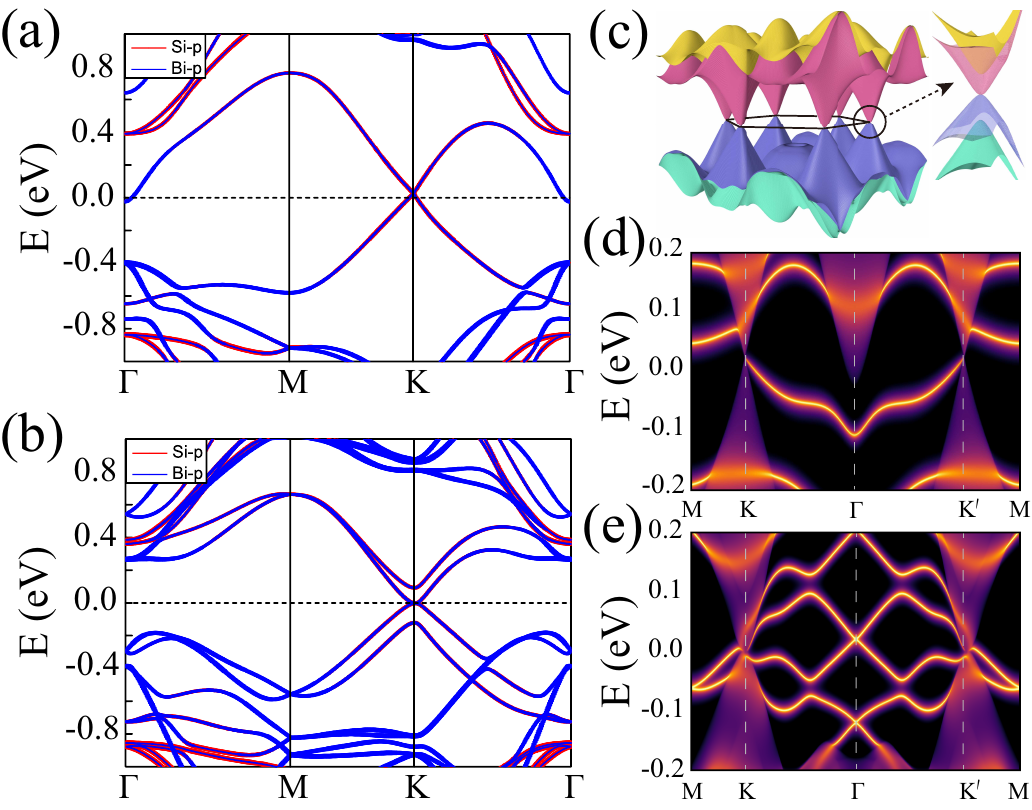}
\caption{\label{bands} The electronic band structure of the Si/Bi heterostructure along the symmetric axes (a) without and (b) with the SOC. (c) The energy dispersions of bands near the Fermi level across the whole Brillouin zone. The corresponding edge states (d) without and (e) with the SOC.}
\end{figure}

\section{Electronic and topological properties}

As is well known, monolayer Si is a Dirac semimetal, while single bilayer Bi is a quantum spin Hall insulator \cite{BiQSH}. When stacking the Si and Bi layers forms a heterostructure, the reduced symmetry combined with the enhanced SOC in Si layer due to the proximity effect by the adjacent Bi layer may result in the emergence of quadratic double Weyl phase. Based on this idea, we constructed the Si/Bi heterostructure (Fig. \ref{crystal}) and calculated its electronic band structure along high-symmetry directions, as shown in Fig. \ref{bands}. When ignoring the SOC, the Si/Bi heterostructure is a semimetal with linear dispersion nodes at the $K$ and $K{'}$ points protected by the $C_{3v}$ symmetry (Fig. \ref{bands}(a)). Due to the lack of space-inversion symmetry but the presence of time-reversal symmetry, the two linear dispersion nodes may carry opposite Berry phases. Indeed, our calculations show that the two linear dispersion nodes at the $K$ and $K{'}$ points have Berry phases $\pi$ and -$\pi$, respectively. Thus, without the SOC, the Si/Bi heterostructure is a conventional Weyl semimetal with two Weyl nodes. Here we remind that, in three dimensions, two Weyl nodes related by the time-reversal symmetry carry the same chiral topological charge. This gives rise to at least four Weyl nodes in order to satisfy no-go theorem \cite{NIELSEN198120, NIELSEN1981173}. In contrast, in two dimensions two Weyl nodes related by the time reversal symmetry carry opposite Berry phases, so the minimum number of Weyl nodes is two, for example, the case of the Si/Bi heterostructure.

Once the SOC is introduced, the point group symmetry of electronic states of the Si/Bi heterostructure changes from $C_{3v}$ point group to $C_{3v}$ double group. The group theory analysis indicates that the two-dimensional irreducible representation of $C_{3v}$ point group reduces to two one-dimensional irreducible representations and one two-dimensional irreducible representation of $C_{3v}$ double point group. Correspondingly, the Weyl node at the $K$ ($K{'}$) point splits into two non-degenerate and one double-degenerate bands (Fig. \ref{bands}(b)). Moreover, for the non-degenerate bands there is a large splitting which is more than two orders of magnitude larger than that in pristine silicene. This shows that there is a large SOC in the heterostructure.

The close proximity between the silicene and the single bilayer bismuthene allows the electron wavefunctions from both layers to overlap and hybridize sufficiently, endowing the silicene with strong SOC. As shown in Figs. \ref{bands}(a) and \ref{bands}(b), there is indeed a strong interlayer hybridization between the Si-$p$ and Bi-$p$ orbitals around the Fermi level, much enhancing the SOC in the Si/Bi heterostructure. As a result, around the Fermi energy, the conduction band at the $\Gamma$ point is pushed away and then there are only the double-degenerate bands which cross with each other at the $K$ or $K{'}$ point with a quadratic dispersion, as shown in the energy band structure over the whole Brillouin zone in Fig. \ref{bands}(c). Moreover, the calculated Berry phase around the $K$ ($K{'}$) point is $2\pi$ (-$2\pi$). Thus the Si/Bi heterostructure is a perfect quadratic double Weyl semimetal.

Just like the case of a three-dimensional conventional Weyl semimetal, there are also edge states connecting two Weyl points with opposite Berry curvatures in a two-dimensional conventional Weyl semimetal. In the Si/Bi heterostructure, in the case of ignoring the SOC, there is indeed an edge state linking the two conventional Weyl points at the $K$ and $K{'}$ points (Fig. \ref{bands}(d)). Are two-dimensional and three-dimensional quadratic Weyl semimetals still similar? The answer is yes. In fact, the two-dimensional double Weyl semimetal Si/Bi heterostructure is a zero-gap quantum spin Hall insulator, which has topologically protected helical edge states at the boundary. Thus, there are two edge states connecting the two quadratic double Weyl points at K and K$'$, and the two edge states form a Dirac point at the $\Gamma$ point due to the time-reversal symmetry (Fig. \ref{bands}(e)). Moreover, since the $K$ and $K{'}$ points locate at two boundaries of the Brillouin zone, the two edge states cross the whole Brillouin zone. This is a very important property of Weyl semimetals.

On the other hand, the $C_{3v}$ double group symmetry protects the existence of a double-degenerate band, but unprotects a quadratic dispersion. Thus, a quadratic double Weyl point with Berry phase $\pm$2$\pi$ may transform into three Weyl points with Berry phases $\pm\pi$ and one Weyl point with Berry phase $\mp\pi$, even keeping the symmetry of $C_{3v}$ \cite{wu2021unprotected}. However, our calculations show that such a transform is too small to identify in the Si/Bi heterostructure. Hence, it is still rational to consider the crossings at the $K$ ($K{'}$) points as double Weyl points with quadratic band dispersions.

\begin{figure}
\centering
\includegraphics[width=8.5cm]{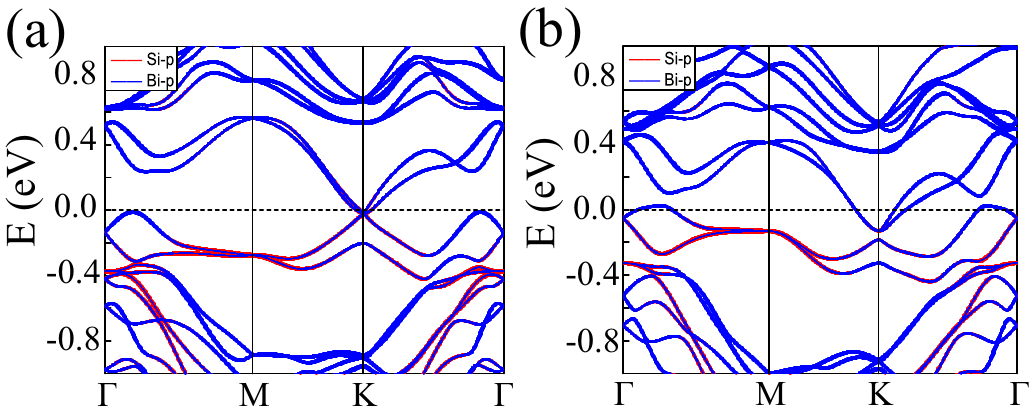}
\caption{\label{strainbands} The orbital-resolved band-structure of Si/Bi heterostructures with the SOC, under compressive biaxial strain of (a) 6.64$\%$, and (b) 8.0$\%$, respectively. The red and blue lines indicate that the states are mainly contributed by the Si-$p$ and Bi-$p$ orbitals, respectively. The Fermi level is set to zero.}
\end{figure}

To further examine the stability of the quadratic double Weyl nodes, we have calculated the electronic band structures of the Si/Bi heterostructure under different strains. The results indicate that when the compressive strain is less than 6.64$\%$, the quadratic double Weyl points always exist. Therefore, the quadratic double Weyl points are stable under a wide range of strain. On the other hand, the quadratic double Weyl semimetal phase transforms into a triple degenerate semimetal with 6.64$\%$ compressive strain applied (Fig. \ref{strainbands}(a)). Since in two dimensions the point groups are uniaxisal and the highest point group symmetry is $D_{6h}$, a triple degenerate semimetal can never be protected by symmetry \cite{PhysRevLett.127.176401}. With the compressive strain further increased, the triple degenerate semimetal then transforms into a trivial semimetal phase, as shown in Fig. \ref{strainbands}(b). The two-dimensional triple degenerate semimetal thus serves as the phase boundary between the quadratic double Weyl phase and trivial semimetal phase.

\begin{figure}
\centering
\includegraphics[width=8cm]{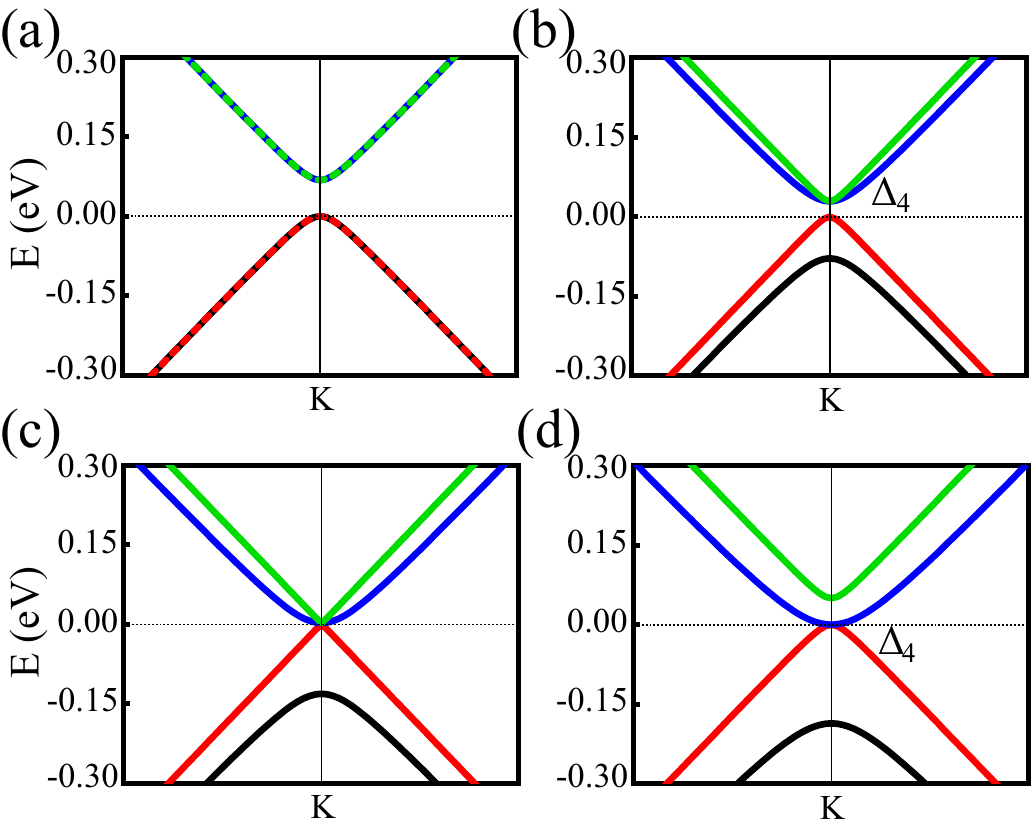}
\caption{\label{TBmodel} The band structures of the Kane-Mele tight-binding model: (a) $\alpha$=0.0, (b) $\alpha$=2.0, (c) $\alpha$=3.33, and (d) $\alpha$=6.0.}
\end{figure}

\section{Kane-Mele tight-binding model}

The quadratic double Weyl nodes are unprotected by the $C_{3v}$ double group symmetry, then, what is the underlying mechanism for the formation of quadratic double Weyl nodes in the Si/Bi heterostructure? To answer the question, we have constructed a low-energy effective model to describe the band structure of the Si/Bi heterostructure, that is, the Kane-Mele tight-binding model \cite{kane2005z}. The effective Hamiltonian is written as
\begin{eqnarray}
\label{Hkm}
H && = \sum_{b=1,2} \{ t_b\sum_{\langle ij\rangle_b} c_i^\dagger c_j
+  i\lambda_{R_b}\sum_{\langle ij\rangle_b} c_i^\dagger
({\bf s}\times\hat{\bf d}_{ij})_z c_j \}  \nonumber \\
&& + \sum_{c=1,2,3} i\lambda_{I_c} \sum_{\langle\langle ij \rangle\rangle_c}\nu_{ij} c_i^\dagger s^z c_j
\end{eqnarray}
in which index $b$ ($c$) represents the two (three) different types of hopping integrals between the nearest (second-nearest) neighbors. Note that there are three different types of bonds in the system. However, the lengths of $b$-$2$ and $b$-$3$ are almost the same, and hence, we set them the same for simplicity. The first term in Eq. \ref{Hkm} is the nearest-neighbor hopping term on the honeycomb lattice, in which $t_b$ is the hopping amplitude. Since the silicene is placed on the substrate of Bi which breaks the space-inversion symmetry, an effective electric field is formed perpendicularly after relaxation, resulting in a Rashba SOC coupling term $i\lambda_{R_b}\sum_{\langle ij\rangle_b} c_i^\dagger({\bf s}\times\hat{\bf d}_{ij})_z c_j$, where $\lambda_{R_b}$ is the amplitude. The third term describes the intrinsic SOC with amplitude $\lambda_{I_c}$. Here we define $\alpha$ as the ratio of the intensity of averaged Rashba SOC to that of averaged intrinsic SOC, namely $\alpha$=$\overline{\lambda}_{R}/\overline{\lambda}_{I}$ (There is only little difference in values between the different types of Rashba SOC or intrinsic SOC, so we simplify the description by averaging the different types of these two SOCs). When ignoring the Rashba SOC ($\alpha$=0.0), there are two double-degenerate bands at the $K$ ($K{'}$) point near the Fermi level. A gap opens due to the existence of the intrinsic SOC (Fig. \ref{TBmodel}(a)). When the Rashba SOC is introduced, as $\alpha$ is finite, the lower double-degenerate bands split. The splitting gap gradually increases with the increase of $\alpha$ (Fig. \ref{TBmodel}(b)). When $\alpha$ reaches a critical value, the higher valence band touches with the double-degenerate conduction bands, forming a triple-degenerate point (Fig. \ref{TBmodel}(c)). With the further increase of $\alpha$ which indicates a very strong Rashba SOC, the triple-degenerate point splits into a non-degenerate point and a quadratic double Weyl point (Fig. \ref{TBmodel}(d)). Note that in the Si/Bi heterostructure, the Rashba SOC is already very large in the optimized configuration, and the system is in the quadratic double Weyl regime. Although an external compressive biaxial strain results in an increased Rashba SOC, this also reduces the distance between the lowest Si and highest Bi atoms, strengthening the interlayer hybridizations. The proximity-induced intrinsic SOC in the Si/Bi heterostructure is enhanced much more rapidly than the Rashba SOC. This gives rise to the transition from the quadratic double Weyl phase to the triple-degenerate phase. Therefore, the novel quantum topological phases in the Si/Bi heterostructure are derived from the competition between the Rashba SOC and the proximity-effect-enhanced intrinsic SOC.

\begin{figure}
\centering
\includegraphics[width=8.5cm]{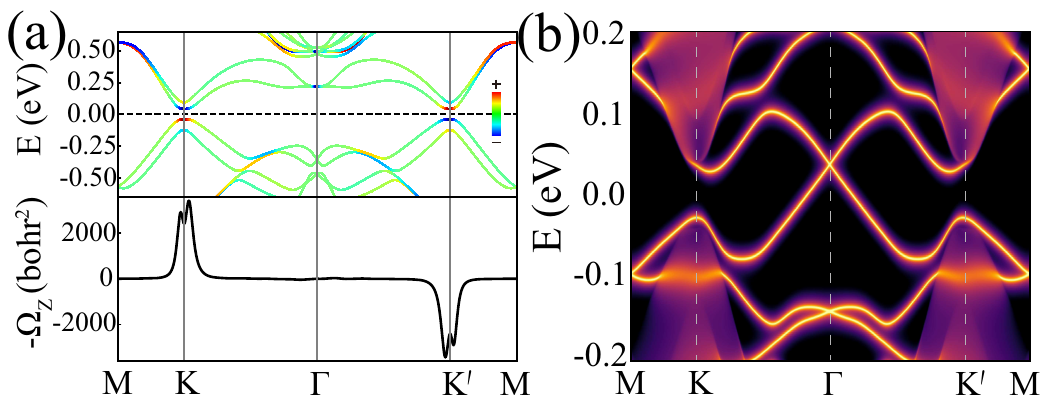}
\caption{\label{NoMirrorbands} (a) The spin-resolved electronic band structure and Berry curvature of the Si/Bi heterostructure with breaking the mirror symmetry. (b) The corresponding edge states.}
\end{figure}

Next, a topological phase transition in the Si/Bi heterostructure is discussed with breaking the $C_{3v}$ symmetry. Specifically, we break the mirror symmetry but retain the $C_{3}$ rotational symmetry. The quadratic double Weyl point at the $K$ ($K{'}$) point then opens a gap and the Si/Bi heterostructure becomes an insulator, as shown in Fig. \ref{NoMirrorbands}(a). Moreover, the calculation of $Z_2$ topological invariant indicates that the system is a quantum spin Hall insulator. The edge states connect the bulk conduction and valence bands, forming a Dirac point at the $\Gamma$ point due to the protection of the time-reversal symmetry, as shown in Fig. \ref{NoMirrorbands}(b). Meanwhile, from Fig. \ref{NoMirrorbands}(a), the distribution of Berry curvature shows that the Berry phases around the $K$ and $K{'}$ points are opposite, 2$\pi$ and \mbox{-2$\pi$}, respectively, which is determined by the time-reversal symmetry. The corresponding Chern numbers are then $C_K$=-$C_{K{'}}$=1, indicating that the Si/Bi heterostructure is a quantum valley Hall insulator \cite{PhysRevX.5.011040,PhysRevLett.99.236809,PhysRevB.84.195444,PhysRevB.81.081403}. In a word, the Si/Bi heterostructure is not only a quantum spin Hall insulator, but also a quantum valley Hall insulator when breaking the $C_{3v}$ symmetry. More importantly, the bilayer Bi was synthesized on the Bi$_2$Te$_3$ substrate and our calculations indicate that a tri-layered Si/Bi/Bi$_2$Te$_3$ heterostructure is still a quadratic double Weyl semimetal. Thus the 2D unconventional Weyl semimetal Si/Bi heterostructure is highly likely to be grown on the Bi$_2$Te$_3$ substrate.

To the end, we would like to remind that such a perfect quadratic double Weyl semimetal is also an ideal platform for hosting topological superconductivity via superconducting proximity effect, for example, setting the Si/Bi heterostructure on a Pb(111) substrate. There has been report that the fabrication of a Pb(111) film on multi-layer Bi enables observation of the proximity-induced superconductivity in Bi(111) as evident from a 1 meV energy gap at 5 K \cite{PbBiHetero}. Together with our findings, the Si/Bi/Pb(111) tri-layered heterostructure could serve as a versatile platform to study the interplay among the proximity-induced superconductivity, electronic, and topological physics.

\section{Conclusion}

In summary, based on the symmetry analysis and the first-principles electronic structure calculations, we propose a promising realistic material system, the Si/Bi heterostructure, to realize the novel two-dimensional quadratic double Weyl state, which arises from the competition between the Rashba SOC and the proximity-effect-enhanced intrinsic SOC. With the application of strain or breaking the mirror symmetry, more exotic topological phases, such as two-dimensional triple degenerate state, quantum spin Hall state, and quantum valley Hall state, can be obtained. Our work therefore provides a good playground to study a variety of novel topological properties, and is of great instructive value for the future work on designing or customizing the electronic and topological properties of materials.

\section*{Conflict of interest}
The authors declare that they have no conflict of interest.

\section*{Acknowledgments}
We thank Z. Yuan for helpful discussions. This work was supported by the National Natural Science Foundation of China under Grants No. 12074040, No. 11934020, and No. 11674027. P.-J. G. was supported by China Postdoctoral Science Foundation funded project (Grant No. 2020TQ0347).

\vspace{1cm}

\bibliographystyle{elsarticle-num-names}
\bibliography{reference}

\end{document}